\documentclass{aastex}

\slugcomment{Draft version, December 2010}

\shorttitle{NEUTRON STAR MASS DETERMINATION}
\shortauthors{RAWLS, M. L. ET AL.}

\begin{document}

\title{Refined Neutron-Star Mass Determinations for 
Six Eclipsing X-Ray Pulsar Binaries$^{\dagger}$}
\altaffiltext{$\dagger$}{This paper includes data gathered with the 6.5 meter Magellan Telescopes located at Las Campanas Observatory, Chile.}


\author{Meredith L. Rawls\altaffilmark{1} and Jerome A. Orosz}
\affil{Department of Astronomy, San Diego State University, 
5500 Campanile Drive, San Diego, CA 92182-1221}
\email{mrawls@sciences.sdsu.edu, orosz@sciences.sdsu.edu}
\altaffiltext{1}{Current address: Department of Astronomy, New Mexico State University, P. O. Box 30001, MSC 4500, Las Cruces, NM 88003-8001.}

\author{Jeffrey E. McClintock and Manuel A. P. Torres}
\affil{Harvard-Smithsonian Center for Astrophysics,
60 Garden Street, Cambridge, MA 02138}
\email{jmcclintock@cfa.harvard.edu, mtorres@cfa.harvard.edu}

\and

\author{Charles D. Bailyn and Michelle M. Buxton}
\affil{Department of Astronomy, Yale University,
P. O. Box 208101, New Haven, CT 06520-8101}
\email{bailyn@astro.yale.edu, michelle.buxton@yale.edu}

\begin{abstract}
We present an improved method for determining the mass of neutron
stars in eclipsing X-ray pulsar binaries and apply the method to six
systems, namely Vela X-1, 4U 1538-52, SMC X-1, LMC X-4, Cen X-3, and
Her X-1. In previous studies to determine neutron star mass, the X-ray
eclipse duration has been approximated analytically by assuming the
companion star is spherical with an effective Roche lobe radius. We
use a numerical code based on Roche geometry with various optimizers
to analyze the published data for these systems, which we supplement
with new spectroscopic and photometric data for 4U 1538-52. This
allows us to model the eclipse duration more accurately and thus
calculate an improved value for the neutron star mass. The derived
neutron star mass also depends on the assumed Roche lobe filling
factor $\beta$ of the companion star, where $\beta = 1$ indicates a
completely filled Roche lobe. In previous work a range of $\beta$
between 0.9 and 1.0 was usually adopted. We use optical ellipsoidal
light-curve data to constrain $\beta$.  We find neutron star masses of
$1.77 \pm 0.08 ~M_{\odot}$ for Vela X-1, $0.87 \pm 0.07 ~M_{\odot}$
for 4U 1538-52 (eccentric orbit), $1.00 \pm 0.10 ~M_{\odot}$ for 4U
1538-52 (circular orbit), $1.04 \pm 0.09 ~M_{\odot}$ for SMC X-1,
$1.29 \pm 0.05 ~M_{\odot}$ for LMC X-4, $1.49 \pm 0.08 ~M_{\odot}$ for
Cen X-3, and $1.07 \pm 0.36 ~M_{\odot}$ for Her X-1. We discuss the
limits of the approximations that were used to derive the earlier mass
determinations, and we comment on the implications our new masses have
for observationally refining the upper and lower bounds of the neutron
star mass distribution.
\end{abstract}



\section{Introduction}

A neutron star is a compact object that is the remnant of a massive
star.  The structure of a neutron star depends on the equation of
state of nuclear matter under extreme conditions, specifically the
relation between pressure and density in the neutron star
interior. For a given equation of state, a mass-radius relation for
the neutron star and a corresponding maximum mass can be derived.
Many such theoretical equations of state exist, ranging from
``soft''---a mass upper limit as low as $1.5 ~M_{\odot}$
\citep{bro94}---to ``stiff''---a higher upper mass limit near $3
~M_{\odot}$ \citep{kal96}.  The accurate measurement of neutron star
masses is therefore important for our understanding of the equation of
state of matter in such high density situations \citep[e.g., see the recent review by][]{kiz10}.

Eclipsing X-ray binary systems where the X-ray source is a pulsar can
be ideal systems for a dynamical determination of the neutron star's
mass.  The orbital period, the semiamplitude of the optical star's
radial velocity curve, the duration of the X-ray eclipse, and the
projected semimajor axis of the pulsar's orbit (measured from the
pulse arrival times) can be used to find the masses of both stars. We
consider six systems where the required measurements have been made:
Vela X-1, 4U 1538-52, SMC X-1, LMC X-4, Cen X-3, and Her X-1.  The
first five systems listed have OB supergiant companion stars, and the
lattermost system has a somewhat less massive companion ($\sim 2
~M_{\odot}$).

In this paper, we present an improved method for determining the mass
of neutron stars in eclipsing X-ray pulsars. In Section
\ref{analytic}, we review the widely used analytic method for neutron
star mass determination and reproduce previous results for the six
systems. In Section \ref{numeric}, we present our numerical code based
on Roche geometry to analyze the published data for the six
systems. In Section \ref{results}, we present our results for each
individual system with the incorporation of optical light curves. In
Section \ref{discuss}, we discuss the implications of our results for
neutron star formation and the equation of state.

\section{Analytic Method}\label{analytic}

\subsection{Basic Equations}\label{basic}

In this section we review the analytic method introduced by \citet{rap83} and \citet{jos84} that is widely used to measure the mass of the neutron star and its optical companion \citep[e.g., see][]{vke95a,vdm07}. To begin, one can write the masses of the optical companion and the X-ray source ($M_{\mathrm{opt}}$ and $M_{\mathrm{X}}$, respectively) in terms of the mass functions:
\begin{equation}\label{mopt}
M_{\mathrm{opt}} = \frac{K_{\mathrm{X}}^3 P(1 - e^2)^{3/2}}{2\pi G \sin^3 i}(1+q)^2
\end{equation}
and
\begin{equation}\label{mx}
M_{\mathrm{X}} = \frac{K_{\mathrm{opt}}^3 P(1 - e^2)^{3/2}}{2\pi G \sin^3 i} \left(1+\frac{1}{q}\right)^2,
\end{equation}
where $K_{\mathrm{X}}$ and $K_{\mathrm{opt}}$ are the semiamplitudes
of the respective radial velocity curves, $P$ is the period of the orbit, $i$ is the inclination of the orbital plane to the line of sight, $e$ is the eccentricity of the orbit, and $q$ is the mass ratio defined as
\begin{equation}\label{q}
q \equiv \frac{M_{\mathrm{X}}}{M_{\mathrm{opt}}} 
= \frac{K_{\mathrm{opt}}}{K_{\mathrm{X}}}.
\end{equation}
The values for $K_X$ and $P$ can be obtained very accurately from X-ray pulse timing measurements (the projected semi-major axis of the pulsar's orbit in light-seconds, $a_X\sin i$, is usually quoted in publications, from which one finds $K_X = 2 \pi c \ a_X \sin i/P$), and optical and/or UV spectra can provide a value for $K_{\mathrm{opt}}$.

Assuming a spherical companion star, the inclination of the system is related to  the eclipse half-angle\footnote{The eclipse half-angle $\theta_e$, or more specifically the semi-eclipse angle of the neutron star, represents half of the eclipse duration.} $\theta_e$, the stellar radius $R$, and the orbital separation $a$ by
\begin{equation}\label{sini-1}
\sin i = \frac{\sqrt{1-(R/a)^2}}{\cos \theta_e}.
\end{equation}
Following the approach in \citet{rap83}, the radius of the companion star is some fraction of the effective Roche lobe radius
\begin{equation}\label{radius}
R = \beta R_L,
\end{equation}
where $R_L$ is the sphere-equivalent Roche lobe radius. We will refer to the fraction $\beta$ as the ``Roche lobe filling factor.'' Combining Equations (\ref{sini-1}) and (\ref{radius}) yields
\begin{equation}\label{sini}
\sin i = \frac{\sqrt{1 - \beta^2 \left({R_L}/{a} 
\right)^2}}{\cos \theta_e}.
\end{equation}

\citet{rap83} provide an approximate expression for $R_L/a$, the ratio of the effective Roche lobe radius and the orbital separation:
\begin{equation}\label{rla}
\frac{R_L}{a} \approx A + B \log q + C \log^2 q,
\end{equation}
where the constants $A$, $B$, and $C$ are
\begin{equation}\label{a}
A = 0.398 - 0.026 \Omega^2 + 0.004 \Omega^3
\end{equation}
\begin{equation}\label{b}
B = -0.264 + 0.052 \Omega^2 - 0.015 \Omega^3
\end{equation}
\begin{equation}\label{c}
C = -0.023 - 0.005 \Omega^2.
\end{equation}
Here, $\Omega$ is the ratio of the rotational frequency of the optical
companion to the orbital frequency of the system.  In other words, it
is a measure of the degree of synchronous rotation, where $\Omega=1$
is defined to be synchronous.  These four expressions give the value
of $R_L$ to an accuracy of about $2\%$ over the ranges of $0\le \Omega
\le 2$ and $0.02\le q \le 1$ \citep{jos84}. If the orbit is eccentric,
the value of $\beta$ is defined at periastron. The star is assumed to
have the same volume over its entire orbit \citep{avn76}, and at any
given phase outside periastron the value of $\beta$ is adjusted
accordingly.  With a given eccentricity $e$, the orbital phase of the
X-ray eclipse is determined by the argument of periastron $\omega$.

For a given system, one can calculate the neutron star mass using
Equations (\ref{mopt}) through (\ref{c}) when given values of $P$,
$a_X\sin i$, $\theta_e$, $K_{\rm opt}$, $\Omega$, and $\beta$ (and if
the orbit is eccentric, $e$ and $\omega$). It is possible to estimate
$\Omega$ by measuring the projected rotational velocity,
$v_{\mathrm{rot}} \sin i$, of the optical companion star. This process
is described in \citet{vdm07} and is employed there for the systems
SMC X-1, LMC X-4, and Cen X-3. Finally, one must assume some value for
the Roche lobe filling factor, $\beta$.  Many of the wind-fed systems
discussed here are thought to be close to filling their Roche lobes,
and a range of $\beta$ between 0.9 and 1.0 is usually adopted
\citep{rap83, vdm07}.  Since all the measured input quantities are not
known exactly, a simple Monte Carlo technique may be used to derive
the most likely values of $M_X$, $M_{\rm opt}$, and $i$, and the
corresponding $1\sigma$ confidence limits.

\begin{figure}
\includegraphics[scale=0.7]{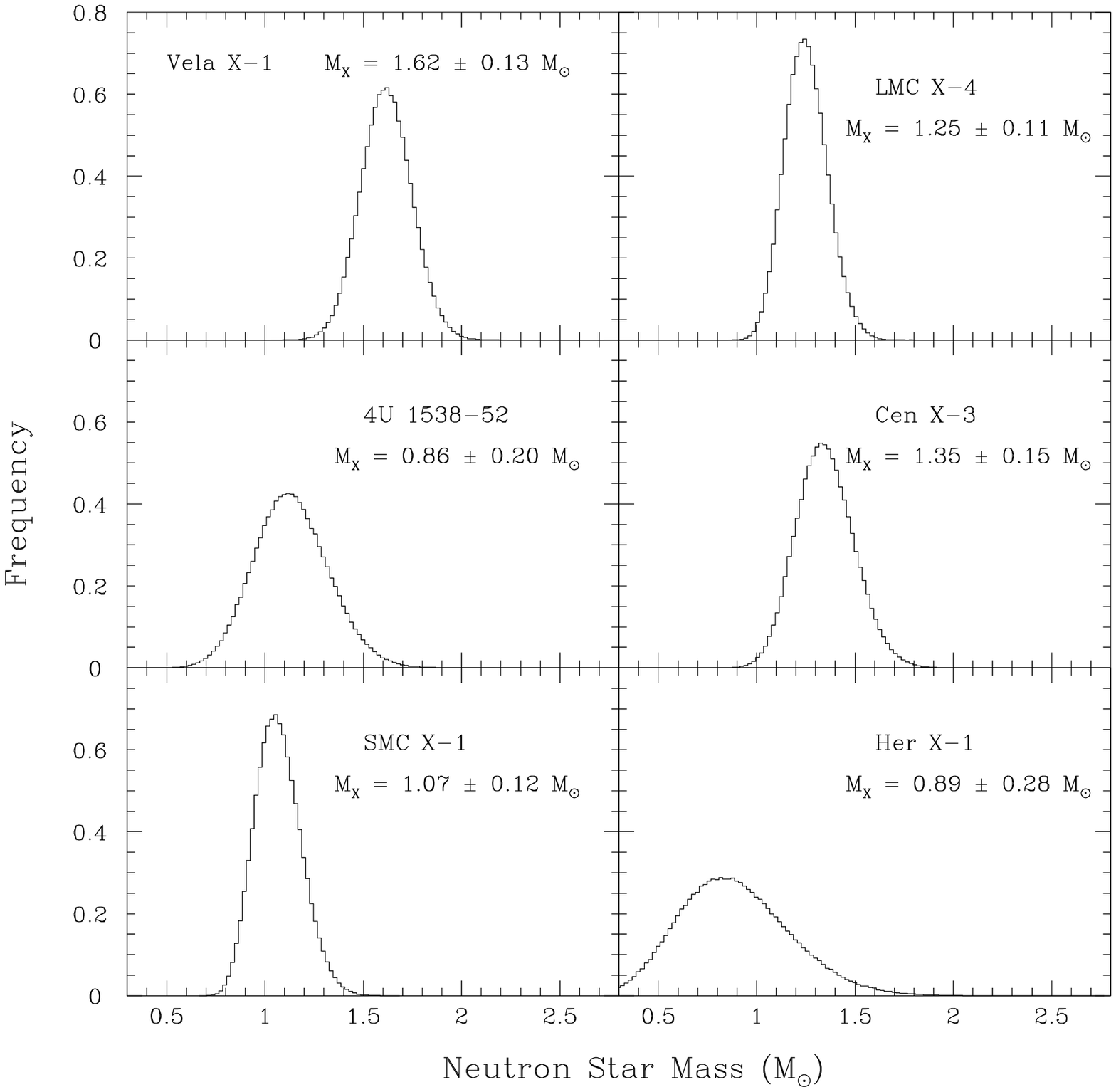}
\caption{Resulting probability distributions (histograms) from Monte Carlo simulations using the analytic method for all six systems as a function of neutron star mass. We assume that any filling factor $0.9 \le \beta \le 1.0$ is equally likely. The mean value $\pm 1\sigma$ is given above each peak. All input parameters for these distributions are given in Tables \ref{input1} and \ref{input2}. We note that our mass for Cen X-3 agree very well with that found by \citet{vdm07}, and the masses for SMC X-1 and LMC X-4 also agree exactly when the values of $\theta_e$ given in \citet{vdm07} are used. We assume $e = 0$ for 4U 1538-52.}\label{histo}
\end{figure}

To determine neutron star masses in this manner for the six eclipsing
systems where all of the necessary quantities are known or estimated
(see Tables \ref{input1} and \ref{input2} and references therein), we
assume that any value within the range $0.9 < \beta < 1.0$ is equally
likely. We then use a Monte Carlo technique to generate a distribution
of resulting neutron star masses, as shown in Figure \ref{histo}. This
provides a clear visual representation of uncertainties through the
shapes of each distribution, and it provides an estimate of the ``most
likely'' mass for each system. The neutron star masses and system
inclinations found in this manner are presented in Table
\ref{anatable}. The masses derived for SMC X-1, LMC X-4, and Cen X-3
agree very well with those cited in \citet{vdm07}.

We note that there is no physical solution for either Vela X-1 or 4U 1538-52 when using this technique with an eccentric orbit, because the quantity in Equation (\ref{sini}) is larger than unity. Adjusting the argument of periastron $\omega$ and/or the eclipse duration $\theta_e$ can force a solution. However, for 4U 1538-52, no solution exists within the $1\sigma$ uncertainties of $\omega$ and $\theta_e$. This discrepancy arises due a high inclination and is discussed further in Sections \ref{Vela} and \ref{4U}. As a workaround, when employing this analytic technique we use a larger eclipse width \citep[$\theta_e = 33 \pm 3^{\circ}$, from][]{vke95} for Vela X-1 and adopt a circular orbit ($e = 0$) as in \citet{cla00} for 4U 1538-52.

\subsection{An Examination of the Approximations}\label{examine}

The analytic method presented in Section \ref{basic} is straightforward, 
and is easy to implement on a computer.  
However, this method relies on two approximations:
\begin{enumerate}
\item[i.] The computation of the effective Roche lobe radius, 
$R_L/a$, from Equations (\ref{rla})--(\ref{c}), and
\item[ii.] The computation of the X-ray eclipse duration, 
$2\theta_e$, from Equation (\ref{sini}).
\end{enumerate}
We use the Eclipsing Light Curve (ELC) code of \citet{oro00}, which is
based on Roche geometry, to test these two approximations, and we
discuss each one in turn.

Strictly speaking, ``Roche geometry'' applies only to binary systems with circular orbits and co-rotating stars.  Numerous authors have presented generalizations of the Roche potential to account for situations in which one or both of these assumption are not met \citep[e.g.,][]{avn75, avn76, wil79, avn83}. These generalizations do not fully describe complete dynamics of the star, and as a result some small approximations  are involved (e.g.\ Wilson 1979). The main assumption is that the timescale for the internal motions of the star that are required for the star to adjust to the varying potential is considerably shorter than the orbital period. Given this, one can compute an effective potential locally at each orbital phase without significant inconsistency \citep{wil79}.  Although these generalized potentials are widely used, it is not known how well they work in practice. For most of the systems discussed here, the orbits are circular and the stars rotate close to the synchronous rate, so the modified potential we use \citep[the ELC code uses the potential given in][]{wil79} should be fairly accurate. Finally, we note that \citet{rap83} also adopted a modified ``Roche potential'' in their analysis---their fitting functions are approximations to numerical integrations of the critical potential surface. Therefore, any systematic error ELC would have owing to improper generalizations of the Roche model would also be present in the work of \citet{rap83} and others that used these fitting functions such as \citet{vdm07}.

\begin{figure}
\includegraphics[scale=0.6,angle=270]{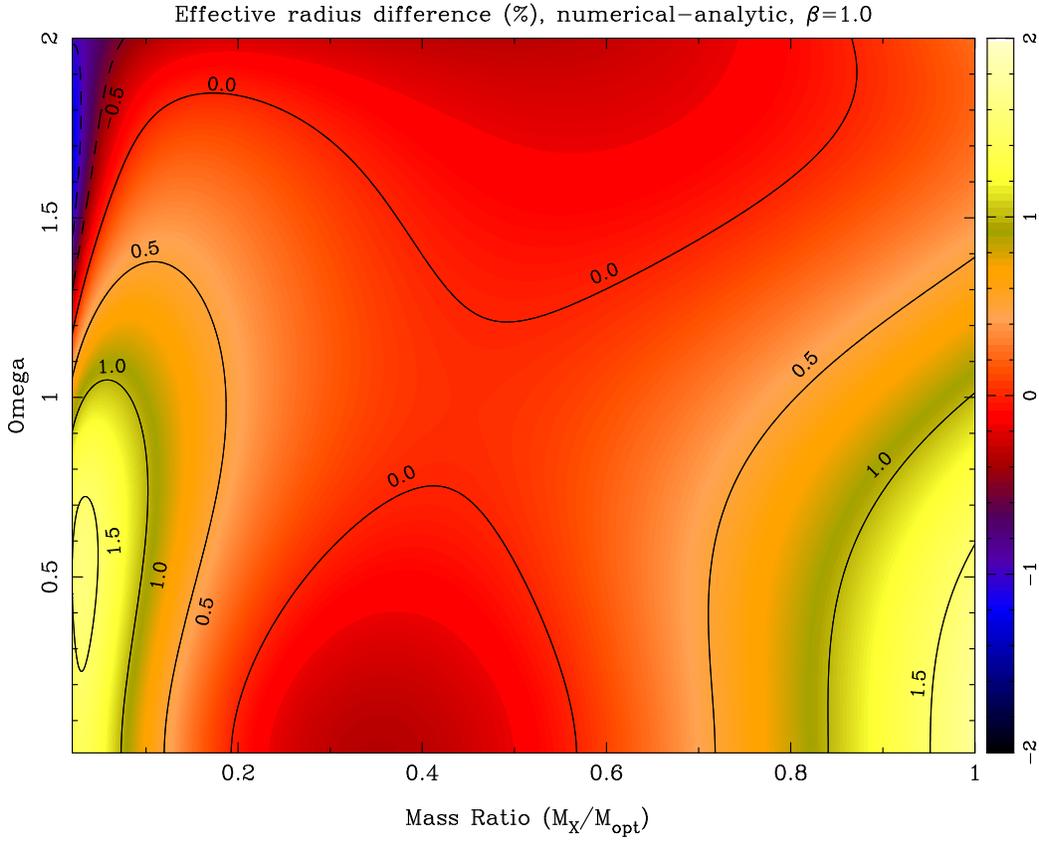}
\caption{Percent difference between the numerically computed effective Roche lobe radius and the analytically computed one (Equations (\ref{rla})--(\ref{c})) as a function of the mass ratio $q$ and the parameter $\Omega$. The largest deviations are about 1.5\% for mass ratios near 0.02 and 1.0.}\label{raddiff}
\end{figure}

The shape and size of the critical potential surface (hereafter the ``Roche lobe'') depend only on the mass ratio $q$ and the parameter $\Omega$. When given $q$ and $\Omega$, it is straightforward to define the equipotential surface (from the value of the gravitational potential at the inner Lagrangian point) and to numerically integrate its volume. The sphere-equivalent radius $R_L$ then follows. We define a large grid of points in the $q$-$\Omega$ plane and compute values of $R_L$, and compare them with the values of $R_L$ found from Equations (\ref{rla}) through (\ref{c}). The results are shown in Figure \ref{raddiff}. \cite{rap83} claim an accuracy of their fitting functions of about $2\%$ over the stated range ($0\le \Omega \le 2$ and $0.02 \le q \le 1$), and our results confirm this.

From Equation (\ref{sini}), one can see that the duration of the X-ray
eclipse depends on the inclination $i$, the Roche lobe filling factor
$\beta$, and $R_L/a$, which is a function of the mass ratio $q$ and
the parameter $\Omega$. The ELC code can be used to compute the
duration of the X-ray eclipse for a given geometry. Rather than using
ray tracing to determine whether a point is eclipsed by the companion
star \citep[e.g., see][]{cha76}, ELC locates the limb of the star to
high accuracy by testing the viewing angles of each surface element.
ELC then uses bisection to find, at each latitude row, the longitude
of the point that has a viewing angle of $\mu=0$. Once found, the
points on the limb define a polygon in sky coordinates.  At that same
phase, the location of the X-ray source (assumed to be a point source)
in sky coordinates is determined.  A simple test is used to determine
if the sky coordinate of the X-ray source is inside or outside the
polygon defined by the horizon of the star.  ELC uses another
bisection routine to find the orbital phase of the X-ray
eclipse ingress to high accuracy. If the orbit is circular, the
eclipse half angle $\theta_e$ is equal to the ingress phase.
If the orbit is eccentric, the X-ray eclipse egress phase is
also computed, and the eclipse half angle is computed from both the ingress
and egress phases.

\begin{figure}
\includegraphics[scale=0.6,angle=270]{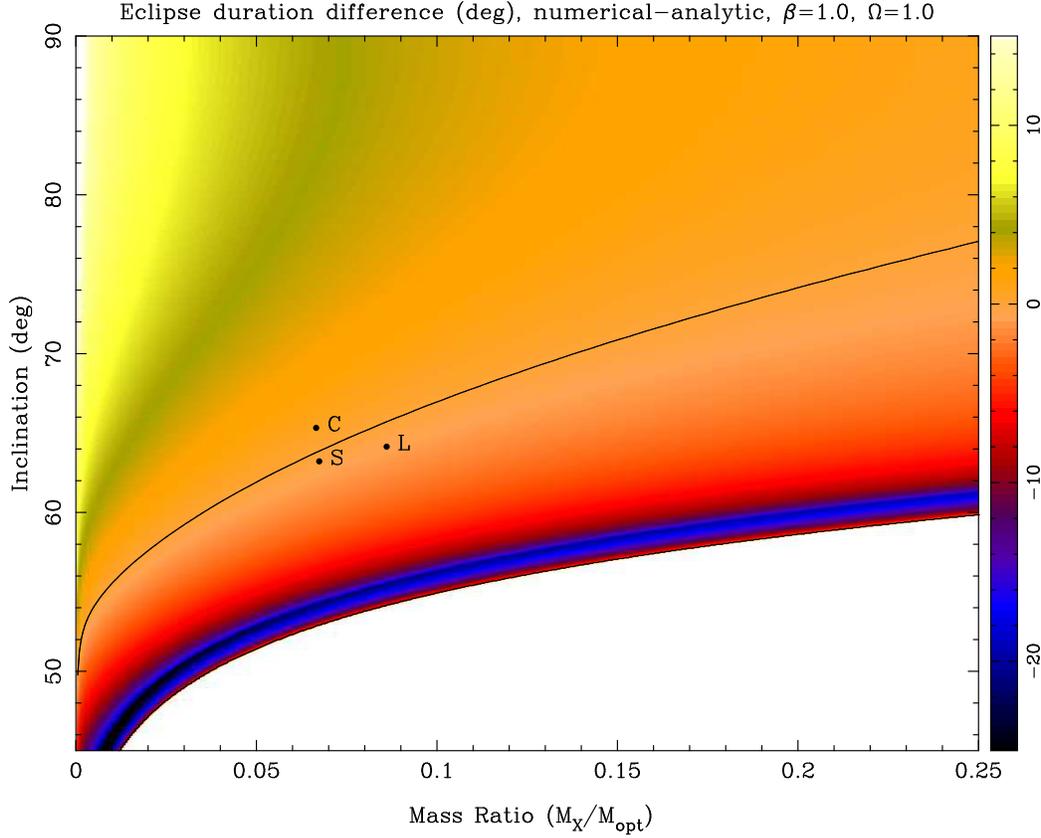}
\caption{Difference (in degrees) of the full X-ray eclipse duration computed numerically and the full X-ray eclipse duration computed analytically via Equations (\ref{sini})--(\ref{c}) as a function of the mass ratio $q$ and the system inclination $i$. Here we set the parameter $\Omega=1$ and the Roche lobe filling factor $\beta=1$. The black curve denotes the contour corresponding to a difference of zero degrees. The locations for three systems are shown:  C=Cen X-3, S=SMC X-1, and L=LMC X-1.}\label{thetadiff1}
\end{figure}

\begin{figure}
\includegraphics[scale=0.6,angle=270]{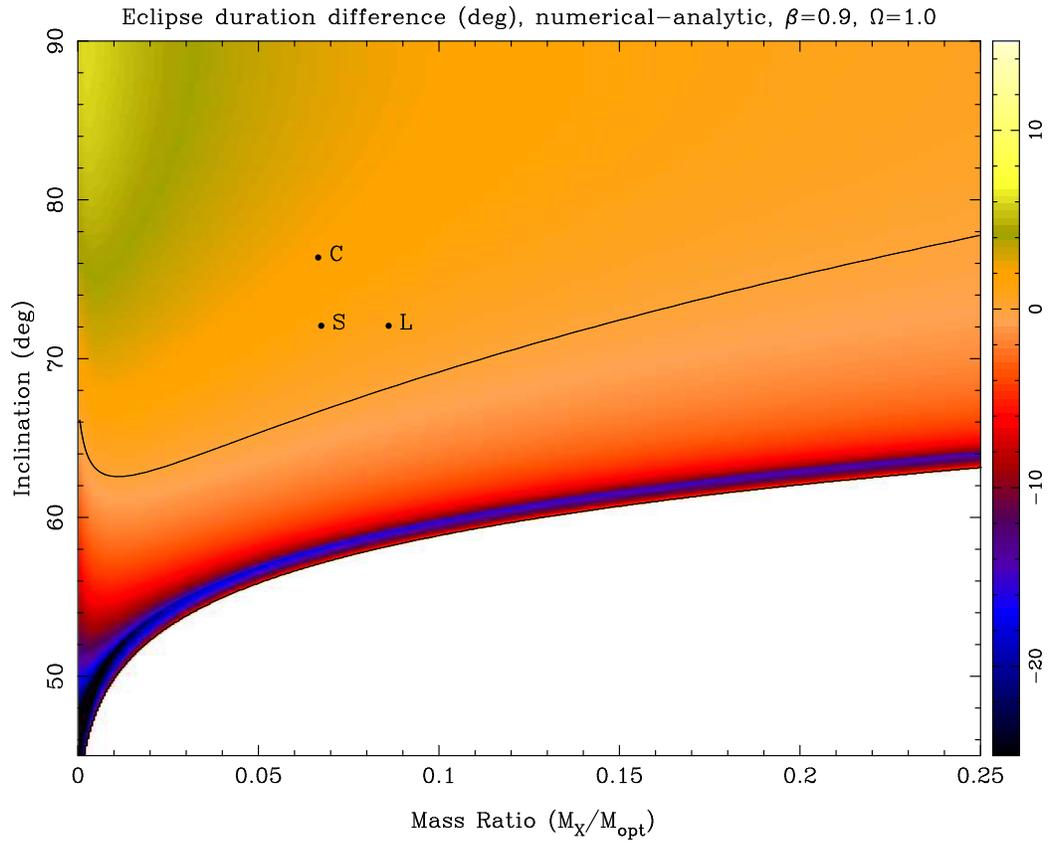}
\caption{The same as in Figure \ref{thetadiff1}, except here the Roche lobe filling factor is $\beta=0.9$.}\label{thetadiff09}
\end{figure}

Setting $\Omega=1$, we use ELC to compute the full duration of the
X-ray eclipse for a wide range of values in the $q$-$i$ plane, using
$\beta=1.0$ and $\beta=0.9$ and assuming a circular orbit.  The full
eclipse duration was also computed from Equation (\ref{sini}), and the
difference between the numerically computed value of $2\theta_e$ and
the analytically computed value of $2\theta_e$ was determined.  Figure
\ref{thetadiff1} shows the differences for $\beta=1$ and Figure
\ref{thetadiff09} shows the differences for $\beta=0.9$. The
differences can be quite extreme: They are in excess of $10^{\circ}$
for small mass ratios and large inclinations and are less than
$-10^{\circ}$ for grazing eclipses.

The approximate locations in the $q$-$i$ plane for three systems (SMC
X-1, LMC X-4, and Cen X-3) are also shown in Figures \ref{thetadiff1}
and \ref{thetadiff09}. Since Her X-1 is a low mass X-ray binary, its
mass ratio does not appear within the limits of the two figures.  Vela
X-1 and 4U 1538-52 are excluded because of their eccentric
orbits. The instantaneous Roche lobe filling factor of an eccentric
system during X-ray eclipse will be less than the value of $\beta$ as
calculated for non-eccentric systems. When $\beta=1$, all three
systems shown in Figures \ref{thetadiff1} and \ref{thetadiff09} are
near the contour denoting zero difference. When $\beta=0.9$, all three
systems are above the zero-difference contour, which indicates that
the eclipse durations computed numerically are {\em longer} than those
computed analytically.

\begin{figure}
\includegraphics[scale=0.8,angle=270]{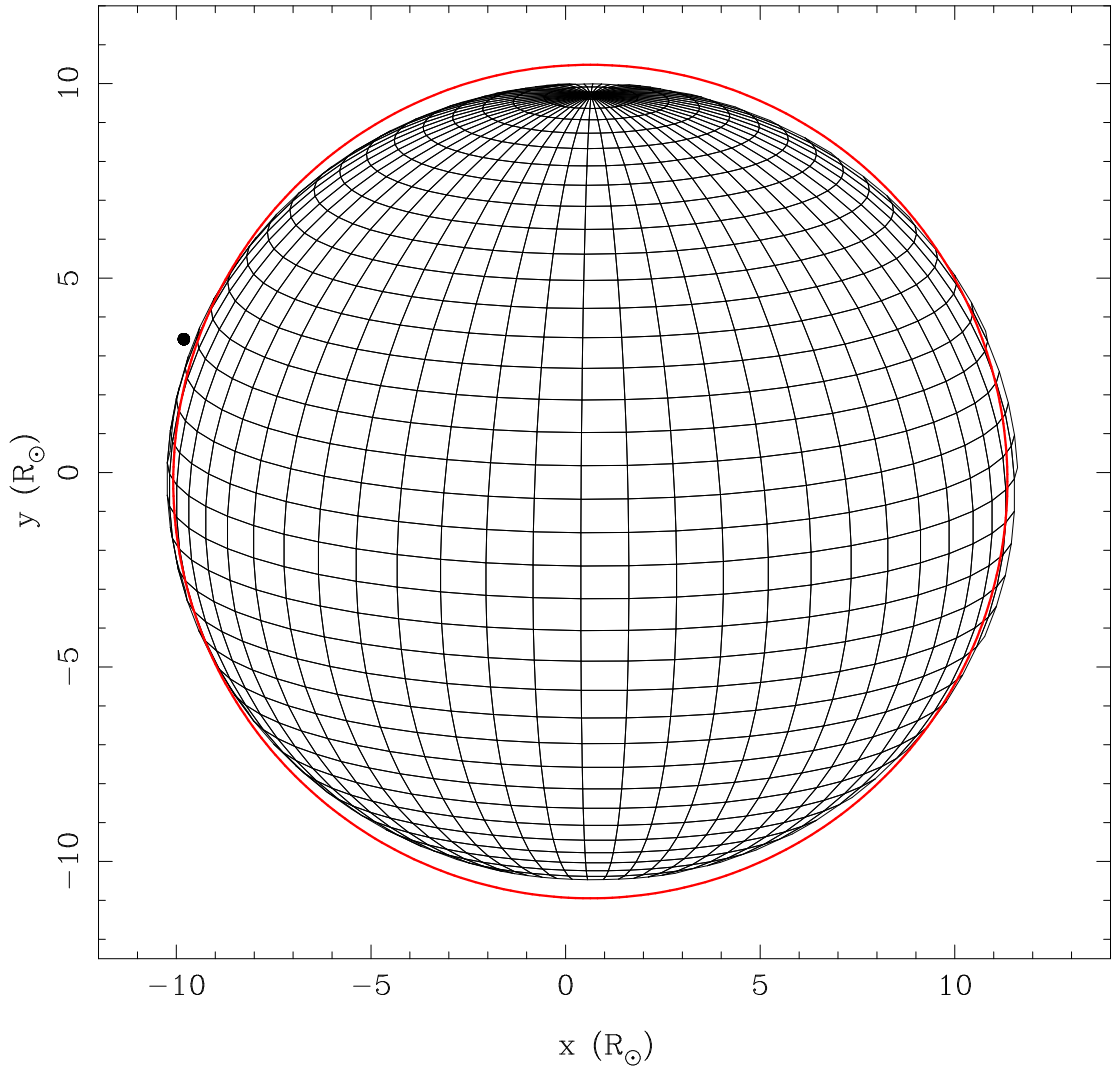}
\caption{Representation in sky coordinates of a system resembling Cen X-3.  The Roche lobe filling factor is $\beta=0.9$, the inclination is $i=76.35^{\circ}$, and the orbital phase is $\phi=34^{\circ}$.  The rotation axis of the companion star and the angular momentum vector of the orbit are parallel to the $y$-axis. The red circle denotes a sphere with the same volume as the companion star. The dot near the ``10 o'clock'' position marks where the neutron star is located at this phase.}\label{schematic1}
\end{figure}

\begin{figure}
\includegraphics[scale=0.8,angle=270]{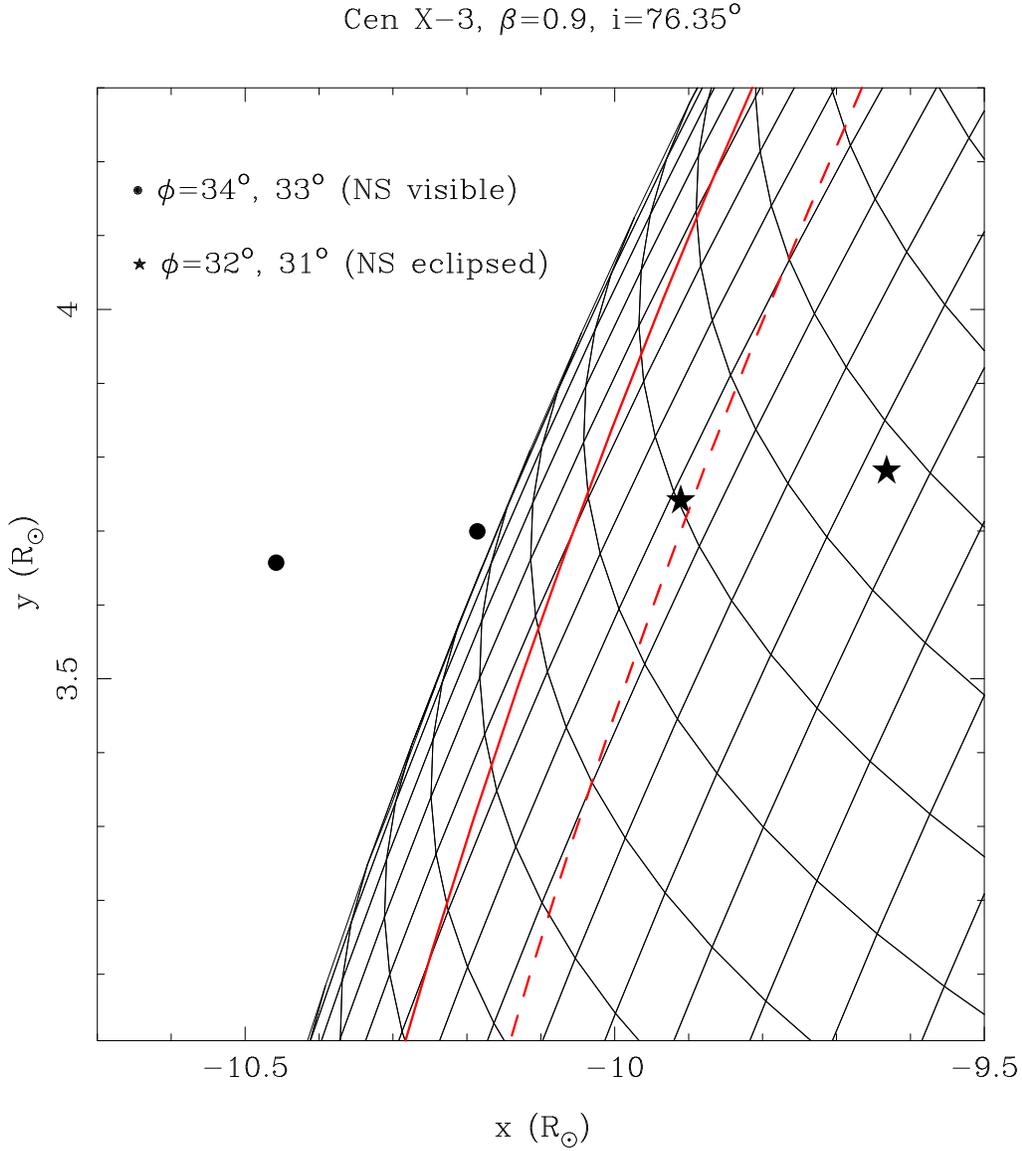}
\caption{Magnified view of the schematic diagram shown in Figure \ref{schematic1}. The solid red line denotes a sphere with the same volume as the star, computed numerically. The dashed line denotes a volume-equivalent sphere computed using the analytic method via Equations (\ref{sini})--(\ref{c}). The location of the neutron star is shown at four orbital phases. In this example, the true egress phase from X-ray eclipse would be very close to $33^{\circ}$ while the egress phase computed using the analytic approximations would be very close to $32^{\circ}$.}\label{schematic2}
\end{figure}

To illustrate the differences between the analytic and ELC results,
Figures \ref{schematic1} and \ref{schematic2} show a system resembling
Cen X-3 in sky coordinates, where the Roche lobe filling factor is
$\beta=0.9$.  Figure \ref{schematic1} demonstrates that the companion
star is not spherical.  The parts of the star near its equator extend
beyond the circle denoting the volume-equivalent sphere.  Hence one
would expect that the duration of an X-ray eclipse for very high
inclinations would be {\em longer} than what one would compute from
the analytic approximations, and a glance at Figures \ref{thetadiff1}
and \ref{thetadiff09} confirms this. In a similar manner, Figures
\ref{thetadiff1} and \ref{thetadiff09} show that for grazing eclipses
(i.e., lower inclinations), the numerically computed durations are
{\em shorter} than the analytic approximations. One can see from
Figure \ref{schematic1} that the polar regions of the companion
star are inside the circle denoting the volume-equivalent sphere,
and as a result there would be inclinations at which the analytic
approximations indicate X-ray eclipses when in fact none occur.

Figure \ref{schematic2} shows a magnified view of Figure
\ref{schematic1} near the neutron star eclipse.  In this example, the
egress phase of the X-ray eclipse is very close to $33^{\circ}$, since
the neutron star is just crossing the limb of the companion star. The
limb of the analytically computed volume-equivalent sphere is well
inside the limb of the star. The egress phase of X-ray eclipse would
be near $32^{\circ}$ when the analytic expressions are used, since the
neutron star is just crossing the limb of the sphere at that
phase. Thus, in this example, the full duration of the X-ray eclipse
computed numerically is a full $2^{\circ}$ longer than the duration
computed analytically.

Having tested both approximations, we conclude that the ELC code does
offer a significantly more accurate representation of the physical
system than the analytic method presented in Section \ref{basic}. The
size of the effective Roche lobe radius as a function of $q$ and
$\Omega$ is relatively well represented by the analytic formulae and
ELC offers only a modest improvement. However, the difference in X-ray
eclipse duration can be quite extreme (as much as $\pm 10^{\circ}$)
and has a direct effect on the calculation of the neutron star mass.

\section{Numerical Method}\label{numeric}

We use the ELC code and its various optimizers to analyze the data
given in Table 1 and optical light curves (see Section \ref{results})
to derive the neutron star masses and their uncertainties. The ELC
code has two advantages here. First, Roche geometry is used (i.e.,\ no
approximations are used to find the effective Roche lobe radius or the
X-ray eclipse duration, as discussed in Section
\ref{examine}). Second, when using ELC, one can make use of any number
of other sources of information about the system, such as optical
light curves. When the geometry is specified, one can compute various observable properties of the system and compare them with the observed values using a $\chi^2$ or similar test. One can then find the family of geometries that best match the observed quantities, and from those geometries the masses and other system parameters follow.

To begin, the orbital period $P$ (which is known to high accuracy) and
the orbital separation $a$ give the total mass of the binary via
Kepler's Third Law.  Specifying the mass ratio $q$ then gives the
component masses.  The shape of the companion star is determined when
the Roche lobe filling factor $\beta$ and the parameter $\Omega$ are
given.  The shape of the orbit, if eccentric, is determined from the
eccentricity $e$ and the argument of periastron $\omega$. Finally,
when the inclination $i$ is given, it is possible to find the
$K$-velocities of the components, the rotational velocity of the
companion star, and the duration of the X-ray eclipse. Thus we
initially have an eight-dimensional parameter space ($P$, $a$, $q$,
$\beta$, $\Omega$, $e$, $\omega$, $i$) to search. However, the search
of the parameter space can be simplified. First, we assume the period
$P$ is known exactly and fix it at the appropriate value for each
system. Likewise, the value of the projected semimajor axis of the
pulsar's orbit $a_X\sin i$ is usually known to high accuracy by
measuring the X-ray pulse arrival times, and is also held fixed. Next,
the parameters $a$ and $q$ can be computed from $a_X\sin i$ and the
$K$-velocity of the companion star:
\begin{eqnarray}
K_X &=& \frac{2\pi c(1-e^2)^{3/2}~a_X\sin i}{P} \\
q   & =& \frac{K_{\rm opt}}{K_X} \\
a    &= &\left(1+ \frac{1}{q} \right)c~a_X. 
\end{eqnarray}
These simplifications give us a six-dimensional
parameter space ($K_{\rm opt}$, $\beta$, $\Omega$, $e$,
$\omega$, $i$) to search.

ELC's various optimizers allow the user to specify a wide range of
values for each parameter as well as sets of data points related to
the physical system, such as light curves or radial velocity
curves. ELC forms random sets of parameters and uses each set to
compute a model. The ``fitness'' of each model is defined using a
$\chi^2$ merit function:
\begin{eqnarray}\label{chi2}
\chi^2 &=& \left(\theta_e({\rm mod}) - \theta_e({\rm obs})\over
           \sigma_{\theta_e}\right)^2 \nonumber \\
&+& \left(v_{\rm rot}\sin i({\rm mod}) - v_{\rm rot}\sin i({\rm obs})\over
           \sigma_{v_{\rm rot}\sin i}\right)^2 \nonumber \\
&+&
\left(K_{\rm opt}({\rm mod}) - K_{\rm opt}({\rm obs})\over
           \sigma_{K_{\rm opt}}\right)^2 \nonumber \\
&+&
\left(e({\rm mod}) - e({\rm obs})\over
           \sigma_{e}\right)^2 \nonumber \\
&+&
\left(\omega({\rm mod}) - \omega({\rm obs})\over
           \sigma_{\omega}\right)^2 .
\end{eqnarray}
Here, the notation (mod) means the quantity computed from the model,
the notation (obs) means the observed quantities, and the notation
$\sigma_{()}$ indicates the $1\sigma$ uncertainty of the observed
quantity. We use the values of $v_{\rm{rot}} \sin i$ and $\omega$
given in Table \ref{input2}. For Her X-1, we do not constrain
$v_{\rm{rot}} \sin i$ but assume synchronous rotation. It might seem
strange to have the terms involving $K_{\rm opt}$, $e$, and $\omega$
in the merit function since they are input parameters for the model,
but our experience has been that the optimizers perform better when
the input parameters are drawn from a uniform distribution.

Once the fitness of a given model is determined, new parameter sets
are constructed using either a Monte Carlo Markov chain optimizer or a
genetic algorithm optimizer. In the latter case, a ``breeding''
technique is used that is based on the ``survival of the fittest''
\citep{oro02}. The probability of ``breeding'' is based on a model's
fitness. Random variations (i.e., ``mutations'') are introduced into a
small fraction of the breeding events, and the process of breeding a
new population and evaluating its members is repeated over many
generations. See \citet{cha95} for a more detailed discussion of
genetic algorithms. In both the Monte Carlo Markov chain and the
genetic algorithm, the fitness of each new parameter set is determined
and the process is repeated until convergence is achieved.

Ultimately, we compute hundreds of thousands of models. Each model has
an associated value of $\chi^2$ and various derived parameters
including the component masses, system inclination, etc. A lower
$\chi^2$ value indicates higher fitness. In this particular case, the
minimum possible value of $\chi^2$ is zero since the 
there are more linearly independent input parameters
than observed quantities. To define the $1\sigma$ limits,
the family of models where $\chi^2\le 1$ is found and the
distributions of the various parameters of interest are constructed.

We note that the value of $a_X\sin i$ given for 4U 1538-52 has a
relatively large uncertainty.  To account for this, we modified the
genetic code to allow for a range of $a_X\sin i$ values drawn from the
appropriate Gaussian distribution to be used.  Although this
modification hardly made a difference in the output parameter
distributions, our results do account for the uncertainty in $a_X\sin
i$.

\begin{figure}
\includegraphics[scale=0.7]{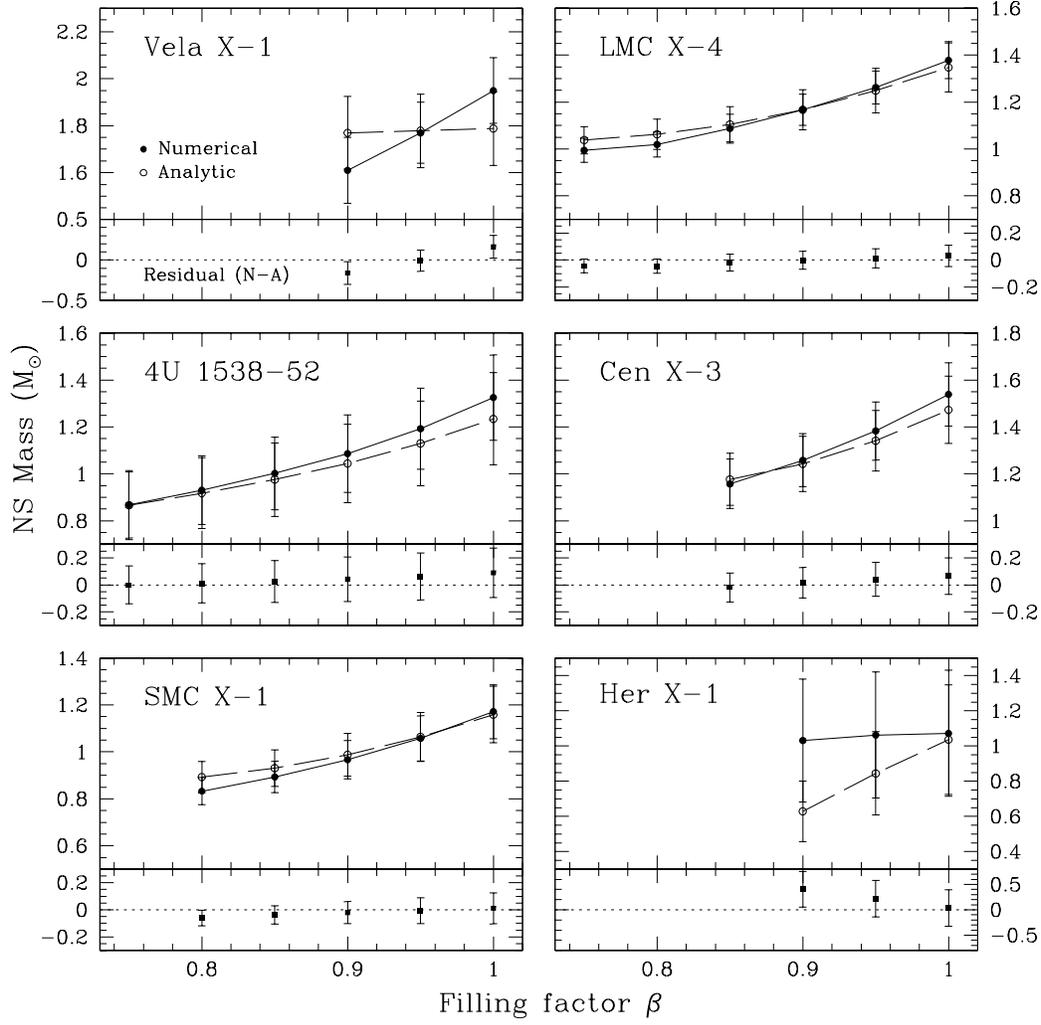}
\caption{Preliminary neutron star mass versus the companion Roche lobe filling factor $\beta$ for all six systems. Analytic indicates masses derived using approximations (see Section \ref{analytic}) while Numerical indicates masses derived using the ELC code based on Roche geometry (see Section \ref{numeric}). Optical light curves have \emph{not} been included in this analysis. Each numerical solution is equally likely in that the eclipse width in the model matches the observed eclipse width exactly. All residuals are in the sense Numerical--Analytic. For simplicity, 4U 1538-52 is approximated as a circular system ($e = 0$).}\label{massdiff}
\end{figure}

We performed a preliminary analysis of the six systems for a range of
$0.75 \le \beta \le 1$ and those results are shown in Figure
\ref{massdiff}. The system inclination is inversely correlated with
$\beta$, and the lower bound on $\beta$ for each system corresponds
roughly to $i = 90^{\circ}$. Hence not all systems can have masses for
all of the values of $\beta$ considered. Two things are apparent in
Figure \ref{massdiff}: First, the neutron star mass is highly
dependent on the choice of $\beta$, and second, the numerical and
analytic results can differ in opposite senses to varying degrees
depending on the choice of $\beta$.

\section{Optical Light Curves}\label{results}

To improve upon the technique described in Section \ref{numeric}, we
consider optical light curves for five systems in addition to the
parameter values in Tables \ref{input1} and \ref{input2}. The shape of
the ellipsoidal light variations from the companion star depends on
the inclination $i$, the mass ratio $q$, the parameter $\Omega$, and
the companion star's Roche lobe filling factor $\beta$. Since the
first three parameters are already well-determined from the width of
the X-ray eclipse and the $K$-velocities of each component, $\beta$
should be quite well constrained by including optical light curves in
the analysis. Such an analysis, which we now present, is trivial to do
using ELC and extremely difficult to do using the analytic approximations.

Adding new observables to the ELC analysis adds terms to the $\chi^2$ merit function as originally shown in Equation (\ref{chi2}).  We effectively have
\begin{equation}\label{chi22}
\chi^2_{\rm{new}} = \chi^2 + \sum\limits_{i=1}^N \left( 
		 \frac{(y_{i, ~\rm{mod}} - y_{i, ~\rm{obs}})}{\sigma_i}\right)^2 ,
\end{equation}
where the final term incorporates a set of $N$ observations defined by
observable quantities $y_i$ (e.g., an optical light curve with $N$
data points). Similar terms may be added for additional sets of
observations (e.g., a radial velocity curve).

\begin{figure}
\includegraphics[scale=0.7]{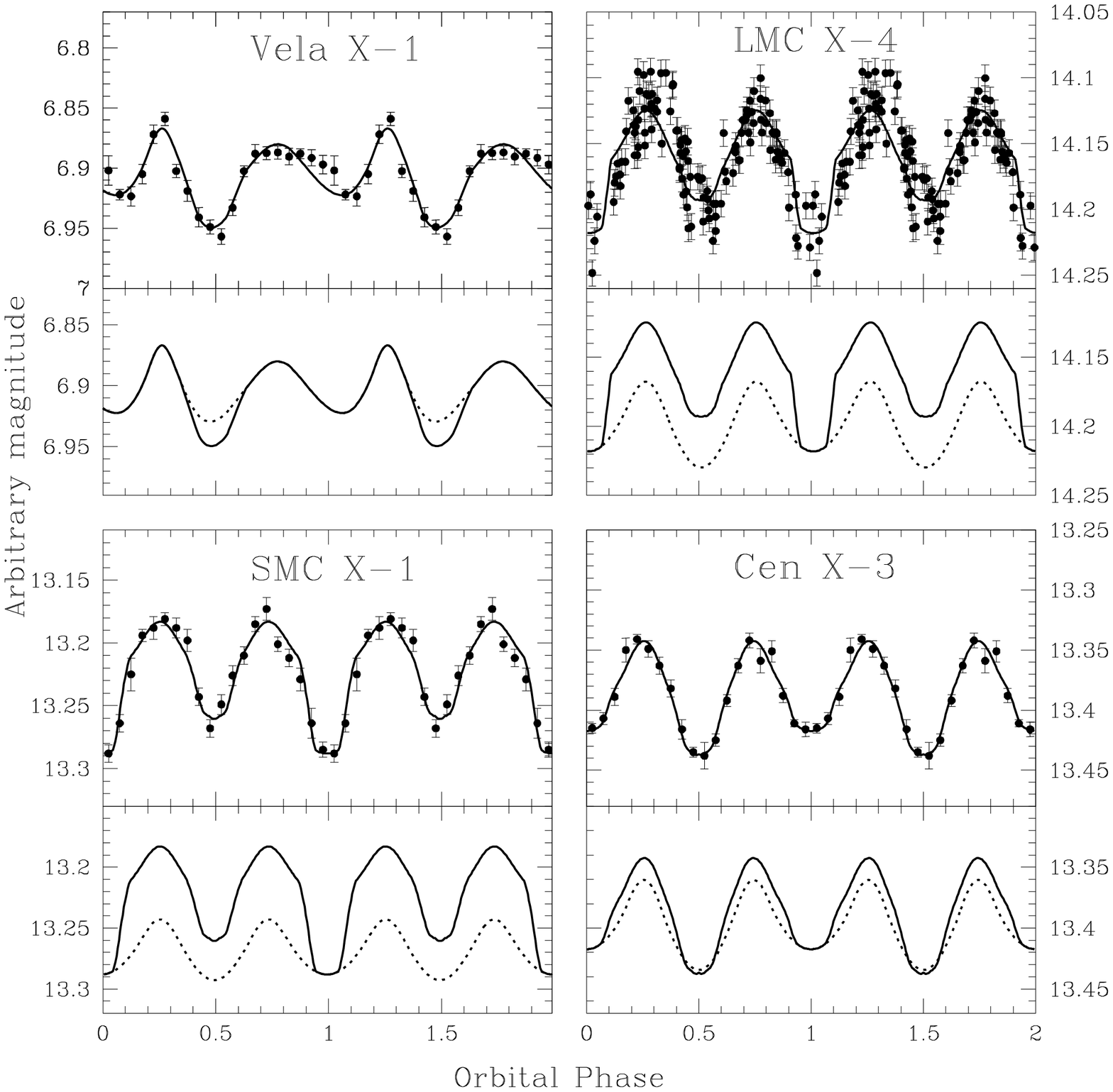}
\caption{Optical light curves for four systems. The solid lines represent the best fit ELC model for each system, all of which include an accretion disk around the neutron star. The dotted lines depict the models with the light from the accretion disk subtracted for comparison. The light curves for Vela X-1, SMC X-1, and Cen X-3 are all V-band data \citep{poj02,pri87,par83}, while the curve for LMC X-4 is B-band data \citep{ilo84}. All data are phased relative to the time of X-ray eclipse.}\label{lightcurves}
\end{figure}

\begin{figure}
\includegraphics[scale=0.7]{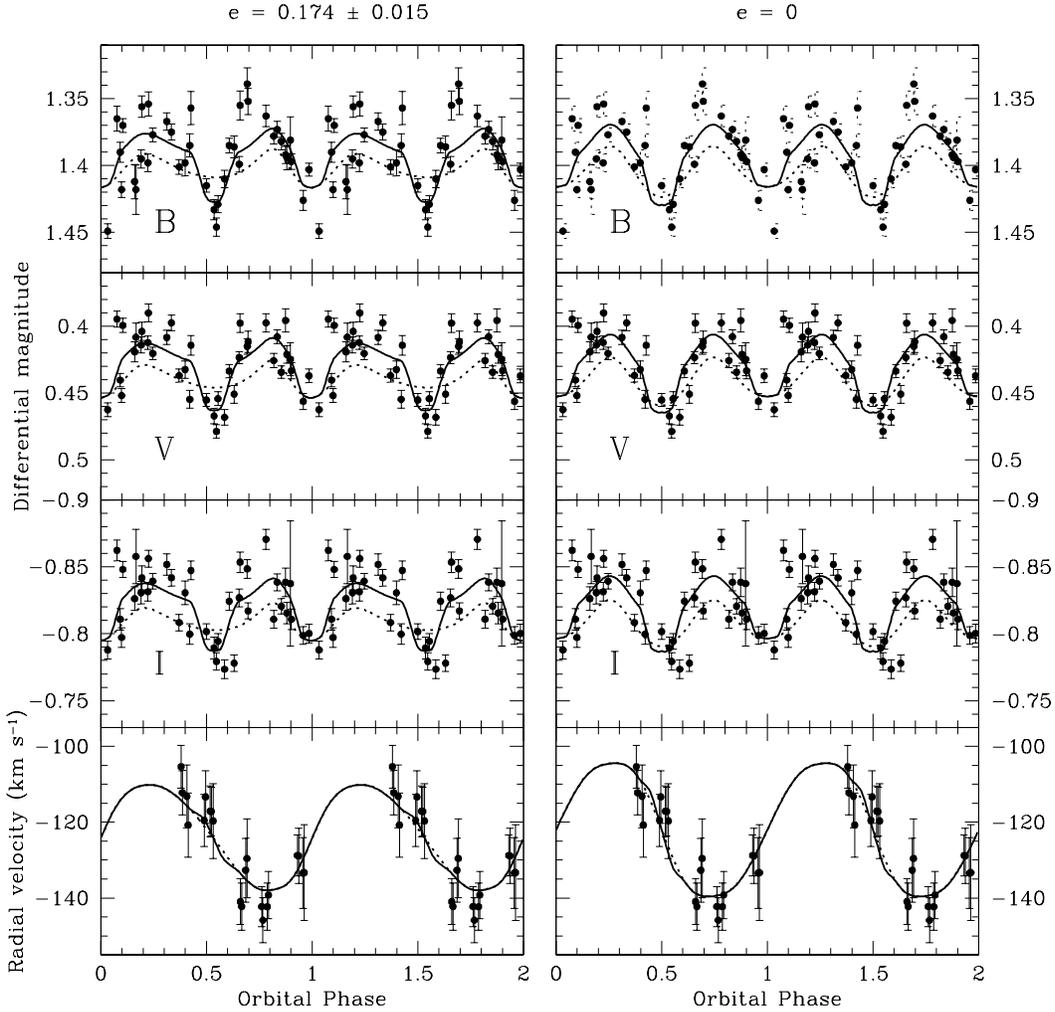}
\caption{Optical \emph{BVI} light curves and radial velocity curves for 4U 1538-52. The solid lines represent the best fit ELC model, which includes an accretion disk around the neutron star. The dotted lines depict the model with the light from the accretion disk subtracted for comparison. The left column shows a model with $e = 0.174 \pm 0.015$ and the right column shows a model for $e = 0$. The eccentric orbit model has a lower overall $\chi^2 = 808$ (versus $\chi^2 = 818$ for the circular orbit), but as discussed in Section \ref{4U} there is no single physical solution for an eccentric orbit and a sufficiently long eclipse duration. All data are phased relative to the time of X-ray eclipse.}\label{4Ufig}
\end{figure}

All of the folded optical light curves discussed below are presented
in Figures \ref{lightcurves} and \ref{4Ufig} with the best fit model
for each system from ELC. For the five systems we analyze in this
manner, the best fit solution includes an accretion disk around the
neutron star. Adding a disk increases the depth of the secondary
eclipse and results in a better fit in all cases. For comparison, we
have plotted both the best fit model and the same model with the light
from the accretion disk subtracted in Figures \ref{lightcurves} and
\ref{4Ufig}. Our final neutron star masses are presented in Table
\ref{tbl-result} and Figure \ref{final} with the corresponding
analytic masses. Rather than using a range of $0.9 \le \beta \le 1$
for the analytic cases, we use the value for $\beta$ returned by the
best fit ELC model for consistency.

\subsection{Vela X-1}\label{Vela}
Vela X-1 has an 8.96 day orbital period and an eccentric orbit with $e
= 0.0898 \pm 0.0012$ \citep{bar01}. The argument of periastron for the
companion star is $\omega = 332.59 \pm 0.92^{\circ}$ \citep{bil97},
where $\omega = \omega_X + 180$. The optical \emph{V} light curve
shown in the upper left panel of Figure \ref{lightcurves} for Vela X-1
is binned data from the All Sky Automated Survey \citep{poj02}. In our
preliminary analysis of Vela X-1 (e.g., Figure \ref{massdiff}), we
adopt a semiduration of the X-ray eclipse $\theta_e = 33 \pm
3^{\circ}$ \citep{vke95}. We later use the more precise value from
\citet{kre08}, $\theta_e = 34.135 \pm 0.5^{\circ}$. Unfortunately, the
analytic technique from Section \ref{analytic} does not arrive at a
physical solution with this longer eclipse duration due to the
system's high inclination. Specifically, solving for the inclination
$i$ is not possible using Equation (\ref{sini}) as $\sin i > 1$ for
all the Monte Carlo simulations. The numerical ELC code does not have
this same limitation. As a workaround, we keep $\theta_e = 33 \pm
3^{\circ}$ for all instances of the analytic case instead. Our final
derived mass for Vela X-1 is $1.77 \pm 0.08 ~M_{\odot}$ with a system
inclination of $77.8 \pm 1.2^{\circ}$.

\subsection{4U 1538-52}\label{4U}
4U 1538-52 has a 3.73 day orbital period and most likely an eccentric
orbit. \citet{cla00} and \citet{muk07} give $e = 0.174 \pm 0.015$ and
$0.18 \pm 0.01$, respectively, which are in agreement. However,
\citet{mak87} give a much lower $e = 0.08 \pm 0.05$ and \citet{vke95}
adopt $e = 0$ in their analysis; even \citet{cla00} provides an
alternate set of fit parameters for a circular orbit. As with Vela
X-1, we find no physical analytic solution for 4U 1538-52 with an
eccentric orbit and its reported eclipse duration ($\theta_e = 28.5
\pm 1.5^{\circ}$) due to the system's high inclination. Therefore, we
follow the approach of \citet{cla00} and present numerical results for
both an eccentric orbit and a circular orbit.

In addition, \citet{cla00} and \citet{muk07} give significantly
different values for $\omega$, the argument of periastron: $244 \pm
9^{\circ}$ and $220 \pm 12^{\circ}$ for the optical source,
respectively. Following the approach of \citet{zha05}, we have
extrapolated a linear decrease of $\omega$ based on the time
difference between the observations made by \citet{cla00} and
\citet{muk07}. We estimate $\omega$ is decreasing by $0.0010 \pm
0.0006^{\circ}$ per day and that our subsequent observations should
therefore have $\omega = 198 \pm 14^{\circ}$. This is the value we
adopt in the numerical model. Even though using a lower $\omega$
nudges the parameter space closer to a single physical analytic
solution, the reported eclipse duration is still too short to allow
$\sin i < 1$ (as in Equation \ref{sini}), and we are forced to retain
a model where $e = 0$ for the analytic case.

Due to the above discrepancies, the reportedly low neutron star mass
($\sim 1 M_{\odot}$) \citep{vke95}, and lack of a published light
curve, we found this system especially worthy of our attention.

Optical light curves in \emph{BVI} for 4U 1538-52 were obtained at the
Cerro Tololo Inter-American Observatory on the 1.3 m SMARTS telescope
with the ANDICAM in June -- September 2009. There are a total of 39
images in each filter taken on different nights, each with a 60-second
exposure time. Standard pipeline reductions were done on all images,
and differential photometry was performed in IRAF for the target and
several comparison stars. Light curves are shown in Figure
\ref{4Ufig}, and observations in all three filters were incorporated
into our numerical analysis.

Twenty-one high resolution spectra of 4U 1538-52 were also taken on
several nights in July and August 2009 at Las Campanas Observatory on
the 6.5 m Clay Magellan telescope with the MIKE spectrograph. Standard
pipeline reductions were done on all frames, including heliocentric
corrections. The spectroscopic analysis was performed in IRAF using
cross-correlation of the blue half of the spectrum (4750 -- 4950 \AA)
with a model B0 star. Radial velocities were then computed and
incorporated into our numerical analysis. The radial velocity curve is
plotted in the bottom panels of Figure \ref{4Ufig} with the adopted
ELC model solution.

We ran two sets of model fits.  We assumed an eccentric orbit with eccentricities in a narrow range around $e=0.18$ and also a circular orbit.
We find $K_{\rm opt}=14.1\pm 1.1$ km s$^{-1}$ for an eccentric orbit and
$K_{\rm opt}=21.8\pm 3.8$ km s$^{-1}$ for a circular orbit.  The
difference between these two measurements is due in part to the noisy 
data and incomplete phase coverage near phase 0.25.
For comparison, \citet{rey92} found $K_{\rm opt}=19.2 \pm 1.2$ km s$^{-1}$ (uncorrected) and $K_{\rm opt}=19.8\pm 1.1$ km s$^{-1}$ (corrected for tidal distortions) assuming a circular orbit.
Our final mass for 4U 1538-52 is quite low: $0.87 \pm 0.07 ~M_{\odot}$
for an eccentric orbit and $1.00 \pm 0.10 ~M_{\odot}$ for a circular
orbit. If the orbit is indeed eccentric, this is an extremely low
neutron star mass.

\subsection{SMC X-1}\label{SMC}
SMC X-1 has a 3.89 day orbital period. It also exhibits a superorbital
X-ray cycle that is probably caused by precession of either the
accretion disk or neutron star \citep[e.g.,][]{pri87}. The optical
\emph{V} light curve for SMC X-1 shown in the lower left panel of
Figure \ref{lightcurves} is from \citet{par84}. Our final derived mass
for SMC X-1 is $1.04 \pm 0.09 ~M_{\odot}$ with a system inclination of
$68.5 \pm 5.2^{\circ}$. We discuss the implications of this low mass result and that of 4U 1538-52 in Section \ref{discuss}.

\subsection{LMC X-4}\label{LMC}
LMC X-4 has a 1.41 day orbital period and, like SMC X-1, also exhibits
a superorbital X-ray cycle. \citet{hee89} performed an extensive study
of the long-term variations in X-ray flux of LMC X-4 and concluded it
contained a warped, precessing accretion disk. The optical \emph{B}
light curve for LMC X-4 shown in the upper right panel of Figure
\ref{lightcurves} is from \citet{ilo84}. Since no data table was
available, the points were extracted from their ``X-ray OFF states''
plot using DEXTER available via the SAO/NASA Astrophysics Data
System. Our final derived mass for LMC X-4 is $1.29 \pm 0.05
~M_{\odot}$ with a system inclination of $67.0 \pm 1.9^{\circ}$.

\subsection{Cen X-3}\label{Cen}
Cen X-3 has a 2.09 day orbital period. The optical \emph{V} light
curve for Cen X-3 shown in the lower right panel of Figure
\ref{lightcurves} is from \citet{par83}. Our final derived mass for
Cen X-3 is $1.49 \pm 0.08 ~M_{\odot}$ with a system inclination of
$66.7 \pm 2.4^{\circ}$.

\subsection{Her X-1}\label{Her}
Her X-1 has a 1.7 day orbital period and a much lower companion star
mass than the other five systems of interest ($\sim 2
M_{\rm{odot}}$). Like SMC X-1 and LMC X-4, Her X-1 also exhibits
superorbital X-ray cycles.

No optical light curve was used for Her X-1 because the optical
$K$-velocity has a very large uncertainty ($90 \pm 20$ km s$^{-1}$)
due to uneven X-ray heating of the companion star. As a result, any
spectral lines measured do not on average emanate from the star's
center of mass. An optical light curve would therefore be ineffective
in constraining the fit. Further, the companion's Roche lobe filling
factor $\beta$ is likely very close to 1 (between 0.95 and 1) as
adopted by and discussed in \citet{vke95} and \citet{bah77}. We
therefore set $\beta = 1$ in our analysis. Our final derived mass for
Her X-1 is $1.07 \pm 0.36 ~M_{\odot}$ with a system inclination
greater than $85.9^{\circ}$.

\section{Discussion}\label{discuss}

\begin{figure}
\includegraphics[scale=0.7]{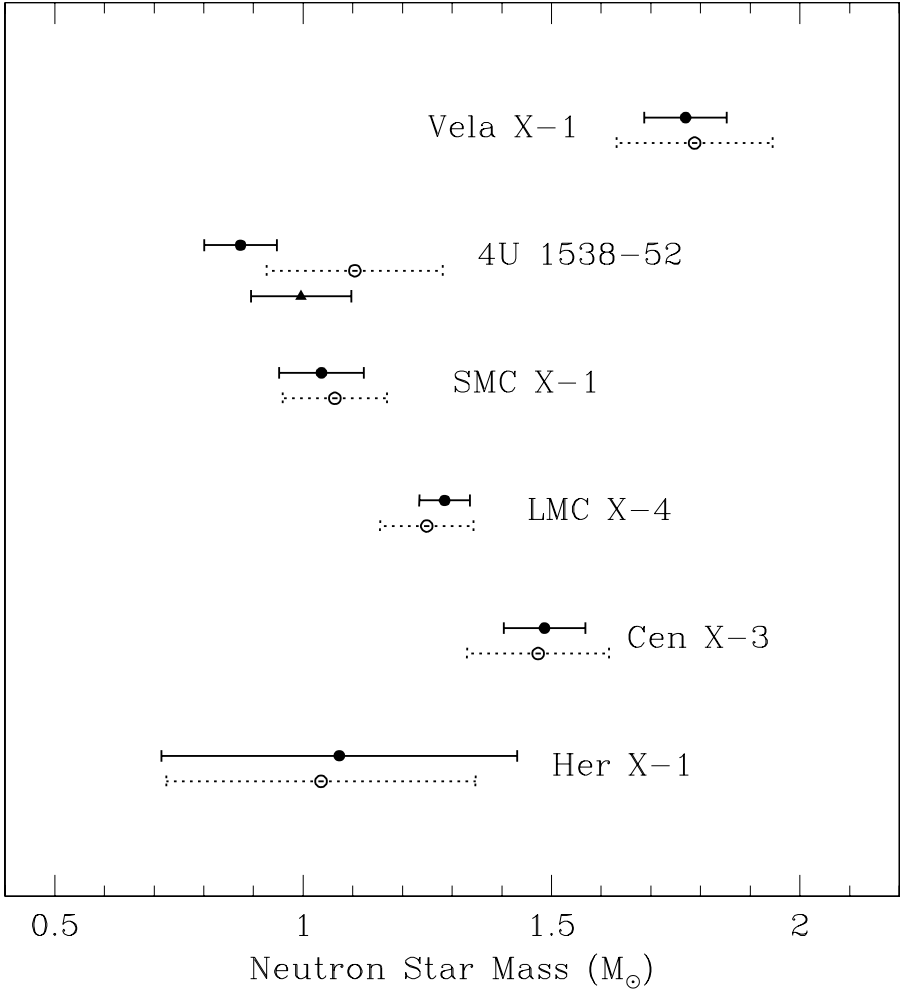}
\caption{Final derived neutron star masses for all six systems. The solid circles represent masses derived using ELC and the open circles are the analytic solutions for comparison. The solid triangle for 4U 1538-52 represents the ELC-derived mass if the system has a circular orbit ($e = 0$), and the open circle for this system also represents a circular orbit (analytic solution). The solid circle for this system is the eccentric orbit mass derived using ELC. All values are given in Table \ref{tbl-result}. For consistency, the fit value of $\beta$ in the ELC models was used in each analytic solution. We assume $\beta = 1$ for Her X-1.}\label{final}
\end{figure}

We present our final derived neutron masses for all six systems in
Figure \ref{final}. These values are also given in Table
\ref{tbl-result}. In every case where we have incorporated light
curves into the analysis, our $1 \sigma$ error bars are smaller than
those of the analytically-derived masses. It is clear from this figure
that these six neutron stars have different masses that span a range
from as low as $0.9 ~M_{\odot}$ to as high as $1.8 ~M_{\odot}$.

\citet{kiz10} recently provided a comprehensive
review of mass determination for neutron stars in double neutron star binaries and in neutron star-white dwarf (NS-WD) binaries. They 
did not include neutron stars in X-ray binaries since the 
mass determinations generally have larger uncertainties than the
typical uncertainties for the double neutron star and NS-WD systems.  The lowest mass neutron star in the \citet{kiz10} sample with a small uncertainty is the companion to PSR J1756-2251 with $M=1.18\pm 0.03\,M_{\odot}$ \citep{fau05}. The largest mass with a small uncertainty is
PSR J1614-2230 $M=1.97 \pm 0.04\,M_{\odot}$ \citep{dem10}.
Other high mass neutron stars include
PSR B1516+02B with $M=2.10\pm 0.19\,M_{\odot}$ \citep{kiz10},
PSR J1748-2446I with $M=1.91^{+0.02}_{-0.10},M_{\odot}$ \citep{kiz10}, and possibly the ``Black  Widow Pulsar'' PSR B1957+20 with
$M=2.40 \pm 0.12,M_{\odot}$ \citep{vke10}.
In the case of the lattermost system, the systematic errors are potentially large owing to the extreme irradiation suffered by the pulsar's evaporating
companion.  The large and secure (low-uncertainty) mass of PSR J1614-2230 rules out many of the so-called soft equations of state \citep{oze10}.

As discussed by \citet{kiz10}, the core mass of a star needs
to exceed the Chandrasekhar mass if it is to end up as a neutron star.
The exact value of the Chandrasekhar mass depends on the electron fraction,
and \citet{kiz10} gives a plausible range of possible neutron star birth masses
of $1.08\lesssim M_{\rm birth}\lesssim 1.57\,M_{\odot}$.  Finally,
as discussed by \citet{kiz10}, the millisecond pulsars in the NS-WD binaries
must have accreted some mass in order to end up with millisecond
spin periods.  They give a range of $0.10\lesssim M_{\rm acc}\lesssim
0.20\,M_{\odot}$ based on angular momentum considerations and on plausible mass transfer rates in the X-ray
binary phase. They conclude that neutron stars with masses less than about $1.1\,M_{\odot}$ would be unusual since it would be difficult to exceed the Chandrasekhar mass, while neutron stars with masses above about $1.8\,M_{\odot}$ must have had either a prolonged stage of mass transfer at an unusually high rate or an unusually high mass when they formed.

Until recently, the neutron star in Vela X-1 has been at the high end of neutron star mass measurements.  Unfortunately, the systematic
errors in the mass determination are difficult to minimize owing
to the non-radial pulsations in the B-giant companion. 
Using the $K$-velocity for Vela X-1 given in \citet{bar01},
the mass of its neutron star is $1.77 \pm 0.08 ~M_{\odot}$ and the
companion star fills its Roche lobe at periastron. \citet{qua03}
derive a slightly higher $K$-velocity for Vela X-1 \citep[$22.6\pm
1.5$ km s$^{-1}$ compared to $21.7\pm 1.6$ km s$^{-1}$
from][]{bar01}, and consequently they derive a higher mass for the
neutron star using the analytic approximations ($2.27 \pm 0.17
~M_{\odot}$ for $\beta=1$). When we used this higher $K$-velocity in
the ELC code, the best fit gave a more conservative neutron star mass
of $1.84 \pm 0.06 ~M_{\odot}$. However, the overall $\chi^2$ was a bit
worse ($\chi^2 = 30.96$ compared to 29.52), and we note that this mass
does barely fall within the $1\sigma$ uncertainty of our adopted
value. In spite of the uncertainties, it appears likely that
the neutron star in Vela X-1 has a relatively 
high mass that is comparable to the masses found for 
PSR B1516+02B, PSR J1748-2446I, and the Black Widow Pulsar.  It is interesting to note that the companion star in Vela X-1 is high mass and will most
likely also produce a neutron star. If so, then Vela X-1 cannot end up in a 
state similar to the Black Widow Pulsar or to the NS-WD binaries 
PSR B1516+02B and PSR J1748-2446I.

On the lower mass end, we have 4U 1538-52 at $0.87 \pm 0.07
~M_{\odot}$ (eccentric) or $1.00 \pm 0.10 ~M_{\odot}$ (circular) and
SMC X-1 at $1.04 \pm 0.09 ~M_{\odot}$. Further spectroscopic
observations are needed to
better establish the shape of the orbit in 4U 1538-52 
and to better establish the $K$-velocity.
In the case of SMC X-1, the observational uncertainties are much
smaller.  Its mass is about 
$1.5\sigma$ smaller than the companion to PSR J1756-2251, which, as
noted above, is the least massive neutron star known with  a
secure mass determination.  The pulse timing properties of SMC X-1 
are very well known, and the only other main quantities that could potentially
change are the $K$-velocity of the companion star and the duration of
the X-ray eclipse. To see what has to change in order to have a
neutron star mass of $\sim 1.2\,M_{\odot}$ in SMC X-1,
we performed a series of fits to the light curve where the $K$-velocity
and eclipse duration were fixed at various values.  To drive the neutron
star mass to higher than $1.2\,M_{\odot}$, the optical $K$-velocity needs to be higher 
($K_{\rm{opt}}\gtrsim 22$ km s$^{-1}$ or almost $2\sigma$ higher than observed) and the eclipse duration needs to be shorter
($\theta_e\lesssim 27.5^{\circ}$ or about $1\sigma$ smaller than observed). These simulations are presented in Table \ref{tblsmc}.
If confirmed, the low masses for the neutron stars in SMC X-1 and
4U 1538-52 would indeed challenge neutron star formation models.

There are a few ways to modestly improve the accuracy of our neutron
star mass determinations.  Since the mass of these neutron stars is
proportional to $K_{\rm opt}^3$, improvements in the measured velocity
curves will improve the mass determinations. For three of the systems
(SMC X-1, LMC X-4, and Cen X-3), this would involve the acquisition
and analysis of a significant number of additional high resolution
optical spectra, work which we did perform in the case of 4U 1538-52
(see Section \ref{4U}). Improvements in the $K$-velocity for Vela X-1 will
be difficult owing to the presence of non-radial pulsations in its B-giant companion \citep{bar01,qua03}. Likewise, improvements in the
$K$-velocity of Her X-1 will be difficult owing to the relatively
strong X-ray heating of its low-mass companion
\citep[e.g.,][]{cra74,rey97}. Modest improvements in the
measured eclipse widths and X-ray timing properties for many of the
systems could be made by observing the sources with greater time
coverage.

\acknowledgments

MLR and JAO gratefully acknowledge the support of NSF grant
AST-0808145. We also thank the anonymous referee who provided
constructive feedback and encouraged us to include the full optical
light curve analysis in this work.

\clearpage

\clearpage

\clearpage

\clearpage

\clearpage

\clearpage

\clearpage

\clearpage

\clearpage

\clearpage

\clearpage

\begin{deluxetable}{lrrrrrrr}
\tabletypesize{\scriptsize}
\tablecolumns{8}
\tablewidth{0pt}
\tablecaption{Input Parameters\label{input1}}
\tablehead{
\colhead{System}                & \colhead{$P_{\rm{orb}}$ (days)}  &
\colhead{$a_X \sin i$ (lt-s)}   & \colhead{$e$\tablenotemark{a}}  &
\colhead{$\theta_e$ (deg)}      & \colhead{$K_{\rm{opt}}$ (km s$^{-1}$)}  &
\colhead{$\Omega$}  		  & \colhead{Ref.}}
\startdata
Vela X-1 					& $8.964368 \pm 0.000040\phm{000}$
& $113.89 \pm 0.13\phm{00}$ 	& $0.0898 \pm 0.0012$
& $34.135 \pm 0.5$\tablenotemark{b}		& $21.7 \pm 1.6$
& $0.67 \pm 0.04\phm{0..}$ 	& 1,3,7 \\
4U 1538-52				& $3.728382 \pm 0.000011\phm{000}$
& $56.6 \pm 0.7$\tablenotemark{c}\phm{0} 	& $0.174 \pm 0.015$\tablenotemark{c}
& $28.5 \pm 1.5$			& $20 \pm 3\phm{0.}$
& $0.91 \pm 0.20$\tablenotemark{d}\phm{.} 	& 2,4,6 \\
SMC X-1 					& $3.89229090 \pm 0.00000043\phm{0}$
& $53.4876 \pm 0.0004$ 		& 0\phm{00000..}
& $28.25 \pm 2.25$ 			& $20.2 \pm 1.1$
& $0.91 \pm 0.20\phm{0..}$ 	& 5 \\
LMC X-4 					& $1.40839776\pm 0.00000026\phm{0}$
& $26.343 \pm 0.016\phm{0}$ 	& 0\phm{00000..}
& $27 \pm 2\phm{0.}$		& $35.1 \pm 1.5$
& $0.97 \pm 0.13\phm{0..}$ 	& 5 \\
Cen X-3 					& $2.08713845 \pm 0.00000005\phm{0}$
& $39.56 \pm 0.07\phm{00}$	& 0\phm{00000..}
& $32.9 \pm 1.4\phm{.}$			& $27.5 \pm 2.3$
& $0.75 \pm 0.13\phm{0.}$ 	& 5 \\
Her X-1 					& $1.700167720 \pm 0.000000010$
& $13.1831 \pm 0.0003$		& 0\phm{00000..}
& $24.5 \pm 0.5$			& $90 \pm 20\phm{.}$
& $1.0 \pm 0.15$\tablenotemark{e}\phm{.} 	& 6\\
\enddata
\tablenotetext{a}{Eccentricities less than 0.01 are approximated to be 0.}
\tablenotetext{b}{For reasons discussed in Section \ref{Vela}, we adopt $33 \pm 3$ from \citet{vke95} for the analytic case instead.}
\tablenotetext{c}{As discussed in Sections \ref{basic} and \ref{4U}, we consider both an eccentric orbit and the case $e = 0$ for 4U 1538-52. When $e = 0$, $a_X \sin i = 54.3 \pm 0.6$ \citep{cla00}.}
\tablenotetext{d}{We use the same $\Omega$ here as for SMC X-1 since the systems' $v_{\rm{rot}} \sin i$ values are equal within uncertainty (see Table \ref{input2}).}
\tablenotetext{e}{Here we assume synchronous rotation ($\Omega = 1$).}

\tablerefs{
(1) Barziv et al. 2001; (2) Clark 2000; (3) Kreykenbohm et al. 2008; (4) Mukherjee et al. 2007; (5) van der Meer et al. 2007 (and references therein); (6) van Kerkwijk et al. 1995a; (7) Zuiderwijk 1995.}

\end{deluxetable}

\clearpage

\begin{deluxetable}{lccc}
\tablecolumns{8}
\tablewidth{0pt}
\tablecaption{Additional Input Parameters\label{input2}}
\tablehead{
\colhead{System}	& \colhead{$v_{\rm{rot}} \sin i$ (km s$^{-1}$)}
& \colhead{$\omega$ (degrees)\tablenotemark{a}}	& \colhead{Ref.}}
\startdata
Vela X-1 	   & $116 \pm 6\phm{0}$	& $332.59 \pm 0.92\phm{0}$	& 1,6 \\
4U 1538-52   & $180 \pm 20$	     & $198 \pm 14$\tablenotemark{b} & 2,3,4 \\
SMC X-1	   & $170 \pm 30$		& 	--		& 5 \\
LMC X-4	   & $240 \pm 25$		& 	--		& 5 \\
Cen X-3	   & $200 \pm 40$		& 	--		& 5 \\
Her X-1	   & \nodata			& 	--		&   \\
\enddata
\tablenotetext{a}{$\omega$ is the argument of periastron for the companion star which is defined only for the systems with nonzero eccentricity (see Table \ref{input1}).}
\tablenotetext{b}{This value assumes a constantly changing $\omega$ over time, as discussed in Section \ref{4U}.}
\tablerefs{(1) Bildsten et al. 1997; (2) Clark 2000; (3) Mukherjee et al. 2007; (4) Reynolds et al. 1992; (5) van der Meer et al. 2007; (6) van Kerkwijk et al. 1995a. }

\end{deluxetable}

\clearpage

\begin{deluxetable}{lcc}
\tablecolumns{3}
\tablewidth{0pt}
\tablecaption{Analytic Neutron Star Masses\tablenotemark{a}\label{anatable}}
\tablehead{
\colhead{System}	& \colhead{$M_X~(M_{\odot})$} 
& \colhead{$i$ (deg)} }
\startdata
Vela X-1 	   & $1.617 \pm 0.130$		& 	$85.9 \pm 2.0$	\\
4U 1538-52\tablenotemark{b}   & $0.859 \pm 0.200$	     & 	$67.3 \pm 4.4$	\\
SMC X-1	   & $1.067 \pm 0.116$		& 	$67.8 \pm 4.7$	\\
LMC X-4	   & $1.252 \pm 0.108$		& 	$68.8 \pm 3.9$	\\
Cen X-3	   & $1.349 \pm 0.146$		& 	$72.5 \pm 4.8$	\\
Her X-1	   & $0.890 \pm 0.280$		& 	$82.0 \pm 3.7$	\\
\enddata
\tablenotetext{a}{All values listed are from Monte Carlo simulations that draw from a random distribution of $0.9 \le \beta \le 1$.}
\tablenotetext{b}{Values for 4U 1538-52 assume $e = 0$ in order to arrive at a physical solution as discussed in Sections \ref{basic} and \ref{4U}.}

\end{deluxetable}

\clearpage

\begin{deluxetable}{lrrcrrrrrr}
\tabletypesize{\scriptsize}
\tablecolumns{6}
\tablewidth{0pt}
\tablecaption{Derived System Parameters\label{tbl-result}}
\tablehead{
\colhead{}     &  	\multicolumn{2}{c}{Analytic\tablenotemark{a}} 
			&  	\colhead{}	
			&	\multicolumn{6}{c}{Numerical} \\
\cline{2-3} \cline{5-10} \\
\colhead{}	&	\colhead{$M_X~(M_{\odot})$}
			&	\colhead{$i$ (deg)} 	
			&	\colhead{}
			&	\colhead{$M_X~(M_{\odot})$}				
			&	\colhead{$i$ (deg)} 
			&	\colhead{$M_{\rm{opt}}~(M_{\odot})$}
			&	\colhead{$R_{\rm{opt}}~(R_{\odot})$\tablenotemark{b}}	
			&	\colhead{$f$\tablenotemark{c}}
			&	\colhead{$\beta$}  
}

\startdata

Vela X-1
& $1.788 \pm 0.157$ 	& $83.6 \pm 3.1$	& 
& $1.770 \pm 0.083$ 	& $78.8 \pm 1.2$ 
& $24.00 \pm 0.37$		& $31.82 \pm 0.28$
& $0.99 \pm 0.01$		& $1$ \\

4U 1538-52 (ecc)
& \nodata 			& \nodata			& 
& $0.874 \pm 0.073$ 	& $68.0 \pm 1.4$ 
& $20.72 \pm 2.27$		& $15.72 \pm 0.52$
& $0.88 \pm 0.02$		& $0.95$ \\

4U 1538-52 (circ)
& $1.104 \pm 0.177$ 	& $72.6 \pm 4.2$	& 
& $0.996 \pm 0.101$ 	& $76.8 \pm 6.7$ 
& $14.13 \pm 2.78$		& $12.53 \pm 2.11$
& $0.76 \pm 0.02$		& $0.88$ \\

SMC X-1
& $1.064 \pm 0.105$ 	& $67.8 \pm 4.2$	& 
& $1.037 \pm 0.085$ 	& $68.5 \pm 5.2$ 
& $15.35 \pm 1.53$		& $15.70 \pm 1.36$
& $0.86 \pm 0.07$		& $0.95$ \\

LMC X-4
& $1.249 \pm 0.094$ 	& $68.8 \pm 3.3$	& 
& $1.285 \pm 0.051$ 	& $67.0 \pm 1.9$ 
& $14.96 \pm 0.58$		& $7.76 \pm 0.32$
& $0.86 \pm 0.03$		& $0.95$ \\

Cen X-3
& $1.473 \pm 0.143$ 	& $67.5 \pm 3.2$	&
& $1.486 \pm 0.082$ 	& $66.7 \pm 2.4$ 
& $22.06 \pm 1.37$		& $12.56 \pm 0.56$
& $> 0.96$			& $1$ \\

Her X-1
& $1.036 \pm 0.311$ 	& $80.5 \pm 3.8$	& 
& $1.073 \pm 0.358$ 	& $> 85.9$
& $2.03 \pm 0.37$		& $3.76 \pm 0.54$
& --					& $1$ 
\enddata

\tablenotetext{a}{For consistency, these analytic values are derived using the $\beta$ values returned by the numerical model rather than a distribution of $0.9 \le \beta \le 1$ as in Table \ref{anatable} and Figure \ref{histo}.}
\tablenotetext{b}{The sphere-equivalent radius of the companion star (e.g., see Figure \ref{schematic1}).}
\tablenotetext{c}{The ELC fill factor, defined as the distance from the companion star's center of mass to the point of the star closest to L1. A value of $f$ maps directly to a value for $\beta$, which is the Roche lobe filling factor expressed in terms of the sphere-equivalent volume radius. For the eccentric systems, both $f$ and $\beta$ are defined at periastron.}

\end{deluxetable}

\clearpage

\begin{deluxetable}{lccc}
\tablecolumns{3}
\tablewidth{0pt}
\tablecaption{Varying $K_{\rm{opt}}$ and $\theta_e$ in SMC X-1\tablenotemark{a} \label{tblsmc}}
\tablehead{
\colhead{$M_X~(M_{\odot})$} 
& \colhead{$K_{\rm{opt}}$ (km s$^{-1}$)} & \colhead{$\theta_e$ (deg)} }
\startdata
$1.091 \pm 0.062$ &  18.0 &  55.0 \\                      
$0.989 \pm 0.056$ &  18.0 &  58.0 \\                     
$0.970 \pm 0.092$ &  18.0 &  61.0 \\                      
$0.904 \pm 0.050$ &  18.0 &  64.0 \\                      
$1.103 \pm 0.061$ &  20.0 &  55.0 \\                    
$1.099 \pm 0.070$ &  20.0 &  58.0 \\                      
$1.002 \pm 0.050$ &  20.0 &  61.0 \\                      
$0.944 \pm 0.055$ &  20.0 &  64.0 \\                      
$1.203 \pm 0.064$ &  22.0 &  55.0 \\                      
$1.117 \pm 0.096$ &  22.0 &  58.0 \\                     
$1.058 \pm 0.047$ &  22.0 &  61.0 \\                      
$1.055 \pm 0.048$ &  22.0 &  64.0 \\                    
$1.260 \pm 0.072$ &  24.0 &  55.0 \\                    
$1.161 \pm 0.049$ &  24.0 &  58.0 \\                    
$1.119 \pm 0.067$ &  24.0 &  61.0 \\                    
$1.078 \pm 0.046$ &  24.0 &  64.0
\enddata
\tablenotetext{a}{These $K_{\rm{opt}}$ and $\theta_e$ values have the same uncertainties as the actual $K_{\rm{opt}}$ and $\theta_e$ for SMC X-1, which are given in Table \ref{input1}.}
\end{deluxetable}

\clearpage

\end{document}